\documentclass[useAMS,usenatbib]{mn2e}

\usepackage{graphicx}
\usepackage{amssymb}
\usepackage{color}
\usepackage{booktabs}
\usepackage{aas_macros}

\newcommand{\Teff}{\ensuremath{T_{\mathrm{eff}}}}%              % T_eff
\newcommand{\logg}{\ensuremath{\log g}}%                        % log g

\newcommand{\bv}{\ensuremath{B\!-\!V}}%                         % B-V
\newcommand{\vk}{\ensuremath{V\!-\!K}}%                         % V-K
\newcommand{\avg}[1]{\left< #1 \right>}% Average
\newcommand{\FeH}{\textrm{[Fe/H]}}
\newcommand{\CFe}{\textrm{[C/Fe]}}
\newcommand{\CH}{\textrm{[C/H]}}
\newcommand{\CaFe}{\textrm{[Ca/Fe]}}
\newcommand{\NFe}{\textrm{[N/Fe]}}
\newcommand{\NH}{\textrm{[N/H]}}
\newcommand{\OFe}{\textrm{[O/Fe]}}
\newcommand{\BaFe}{\textrm{[Ba/Fe]}}
\newcommand{\BaH}{\textrm{[Ba/H]}}
\newcommand{\NaFe}{\textrm{[Na/Fe]}}
\newcommand{\ocen}{\textrm{$\omega$~Cen}}
\definecolor{squares}{RGB}{255,48,0}
\definecolor{triangles}{RGB}{0,86,180}
\definecolor{crosses}{RGB}{0,158,115}
\defcitealias{Johnson2010}{J\&P10}
\defcitealias{vanLoon2007}{vL+07}
\defcitealias{Marino2012}{M+2012}
\defcitealias{Simpson2012}{Paper~I}
\defcitealias{Bellini2009a}{B+09}
\newcommand{\paperone}{\citetalias{Simpson2012}}
\newcommand{\vL}{\citetalias{vanLoon2007}}
\newcommand{\jp}{\citetalias{Johnson2010}}
\newcommand{\bo}{\citetalias{Bellini2009a}}
\title[Abundances of GB of $\omega$~Cen]{Spectral matching for abundances of 848 stars of the giant branches of the globular cluster $\omega$~Centauri}
\author[J. D. Simpson and P. L. Cottrell]{Jeffrey. D. Simpson$^{1}$\thanks{E-mail:
jeffrey.simpson@pg.canterbury.ac.nz} and P. L. Cottrell$^{1}$\\
$^{1}$Department of Physics and Astronomy, University of Canterbury, Private Bag 4800, Christchurch, New Zealand}
\begin{document}

\date{\today}

\pagerange{\pageref{firstpage}--\pageref{lastpage}} \pubyear{2012}

\maketitle

\label{firstpage}

\begin{abstract}
We present the  effective temperatures, surface gravities and abundances of iron, carbon and barium of 848 giant branch stars, of which 557 also have well-defined nitrogen abundances, of the globular cluster $\omega$~Centauri. This work used photometric sources and lower resolution spectra for this abundance analysis. Spectral indices were used to estimate the oxygen abundance of the stars, leading to a determination of whether a particular star was oxygen-rich or oxygen-poor.

The 557-star subset was analyzed in the context of evolutionary groups, with four broad groups identified. These groups suggest that there were at least four main four periods of star formation in the cluster. The exact order of these star formation events is not yet understood. %\textbf{with some models predicting the groups formed from iron-poorest to iron-richest, while others suggest the} potential for iron-poorer groups to form after iron-rich groups.

These results compare well with those found at higher resolution and show the value of more extensive lower resolution spectral surveys. They also highlight the need for large samples of stars when working with a complex object like \ocen.
\end{abstract}


\section{Introduction}

\ocen\ is a large star cluster in the southern sky that straddles the line between a globular cluster (GC) and a dwarf galaxy. With a mass of $2.5\times10^6 M_{\sun}$ \citep{vandeVen2006}, it is smaller than the average dwarf galaxy (small dwarf galaxies $\sim20\times10^6 M_{\sun}$, \citealt{Mateo1998}) but has a larger mass than the typical simple stellar population globular cluster (average mass of $1.9\times10^5 M_{\sun}$, \citealt{Mandushev1991}). Its luminosity also sets it apart from other globular clusters, whose sizes are anticorrelated with brightness: small clusters are the brightest intrinsically (see figure 6 of \citet{Federici2007} for example). Instead \ocen\ is one of three clusters to break this trend, with the other two being M54 and NGC~2419. M54 has a metallicity range like \ocen\ but not as large \citep{Carretta2010} and is found within the core of the Sagittarius Dwarf Elliptical Galaxy. The situation with NGC~2419 is not fully understood, due to its large distance ($m-M$ is 6 magnitudes larger than \ocen) and its small apparent size (5 times smaller than \ocen, \citealt{Harris1996}). \citet{Cohen2010} found a spread of $\sim0.2$ dex in the calcium triplet, which they interpreted as being a metallicity range of the cluster. However \citet{Mucciarelli2012} dispute this, suggesting that the spread is not real, instead the result of not taking into account a spread in magnesium. Higher resolution work by \citet{Cohen2012} found a calcium spread but no spread in iron.

The complex chemical makeup of \ocen\ was first observed photometrically as an intrinsic width to the giant branch (GB) in the colour-magnitude diagram \citep{Cannon1973}. Further studies with photometry and spectroscopy have shown the cluster to consist of multiple distinct GBs \citep{Lee1999,Pancino2000}, at least three main sequences (MS) \citep{Anderson1997,Bedin2004} and a very wide turn-off and sub-giant region \citep{Stanford2010,Villanova2007,Villanova2010}. These different sequences are the result of different metallicities \citep[hereafter \paperone]{Freeman1975,Simpson2012}, helium abundances \citep{Norris2004,Piotto2005,Dupree2011} and maybe even CNO abundances \citep[][hereafter \citetalias{Marino2012}]{Marino2012}. The work presented here aims to expand our knowledge of the cluster by increasing tenfold the number of stars for which carbon and nitrogen abundances have been inferred. These stars are placed into broad evolutionary groups that suggest multiple star birth events have taken place in the cluster's youth.

The two extremes of the evolutionary models for \ocen\ are a merger event \citep[e.g.][]{Icke1988} or complete in situ formation \citep[e.g.][]{Smith2000}. A merger event would require a process that radially well-mixed the stars, since there is not evidence for radially offset inhomogeniety in the stars. \citet[][hereafter \jp]{Johnson2010} did find that there are no oxygen-poor stars at large radial distances in the cluster, but this can be explained by an in situ model via core settling of gas that formed these oxygen-poor populations.

Due to the lack of offsets of the stars' distributions, the in situ model is preferred, but the exact details of this are not yet well understood. \citet{Herwig2012} proposed that Galactic plane passages could be involved, stripping gas from the cluster to allow subsequent generations to form gas that is expelled by stars after these plane passages. They suggest this would explain the sodium-oxygen anticorrelation. However, this feature is observed in every globular cluster \citep{Carretta2009}, so plane passages could be involved but maybe are not important to the chemical evolution of the cluster.

\citet{Valcarce2011} propose that mass is the main driver of the chemical evolution of the cluster. \ocen\ is much more massive that other clusters and this would have allowed it to keep gas that would have been lost through supernov\ae\ and other stellar winds in smaller clusters. This meant that it could form the mulitple generations of stars that are observed in the cluster now. \citet{Valcarce2011} created a toy model but does present an intriguing idea that mass is the most important aspect of the cluster formation. However it does not appear to explain the continuous metallicity distribution of the cluster nor neutron-capture abundances.

\begin{table}
\caption{Qualitative description and quantitative values of the four groups, based upon table 6 of \paperone.}
\label{Table:Qualitative}
 \begin{tabular}{lllll}
 \hline
 Group & [Fe/H] & [C/Fe] & [N/Fe] & [Ba/Fe] \\
 \hline
 1 & Low & Low & Low & Low\\
   & $-1.8$ & $-0.4$ & 0.3 & $-0.4$\\
 2 & Intermediate & High & Low & Intermediate\\
   & $-1.7$ &    0.0 & 0.4 & 0.4\\
 3 & Intermediate & Low & High & Intermediate\\
   & $-1.7$ & $-0.5$ & 1.0 & 0.5\\
 4 & High & Low & High & High\\
   & $-1.3$ & $-0.6$ & 1.8 & 0.8\\
\hline
 \end{tabular}
\end{table}

One of the most important pieces in the puzzle of \ocen\ is the helium content of the stars. This first came to light with the work of \citet{Anderson1997,Bedin2004} who first identified the split main sequence of the cluster. \citet{Norris2004} was the first to propose that this could be the result of differing helium abundaces in the stars. This has been confirmed, but only a small sample of stars have had their helium abundance inferred, which limits the conclusions that can be drawn from this. \citet{Dupree2011} found that of their twelve stars, five had helium lines present in their spectra. The average \NaFe\footnote{Following the standard notation, [X/Fe] denotes the logarithmic difference between X to Fe ratio in the star and the Sun.}\ abundance for those stars with no helium present was $\NaFe=0.0\pm0.3$, while for stars with helium present it was $0.4\pm0.1$. This would suggest that our group \#3 from \paperone\ (intermediate metallicity, Na-rich stars; a description of these groups is found in Table \ref{Table:Qualitative}) are helium-rich, while our groups \#1 and \#2 are standard helium. Further work is needed to increase the number of stars with helium inferred, so that they can be linked to other stars through similar abundances and metallicities.

In this work, it is the carbon and nitrogen that we are most interested in determining. In \paperone\ it was shown that the O-rich stars of the O-Na anticorrelation have different carbon abundances, splitting into a carbon-poor group and a carbon-rich group. It was also found that the former group were iron-poor/Ba-poor, and the latter was iron-intermediate/Ba-intermediate.

These stars also differ in their iron and barium abundances. This implies a process that could increase the iron, s-process and carbon abundance, while keeping the ranges of abundances in nitrogen, oxygen and sodium the same. It was also found that for the intermediate metallicity stars of \citetalias{Marino2012}, although anticorrelated in carbon-nitrogen and sodium-oxygen, were indistinguishable in their range of iron and barium abundances. This requires several processes that appear to have the wrong time periods: the process by which s-process elements are created happens in lower-mass stars than that which is thought to form the sodium-oxygen anticorrelations. How to solve this time problem has not be satisfactorialy explained. It could perhaps require an s-process production site that is not low-mass asymptotic giant branch (AGB) stars \citep{DAntona2011}.

\begin{figure}
\includegraphics[width=84mm]{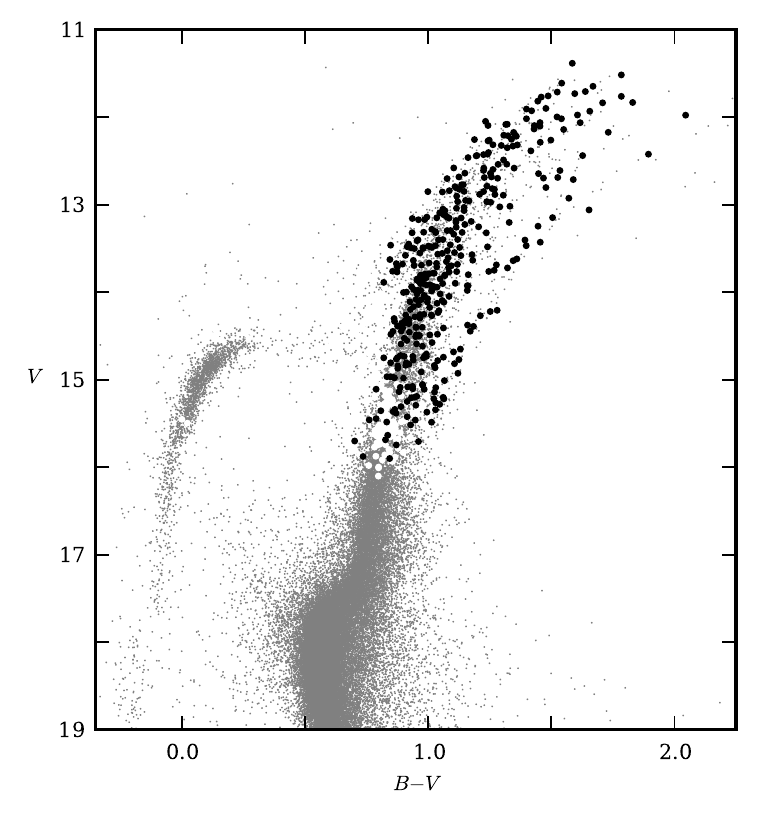}
\caption{Colour-magnitude diagram using the photometry of \bo. The black dots show the 557 stars with $\Teff\leq4900$ K which are discussed in Section \ref{sec:ClusteringAnalysis} and the white-filled circles are some of the remaining 848 stars. The grey dots are stars from \bo. This plot does not show all the stars, as only 65~per~cent of the stars were matched to \bo. \label{fig:CMD_All_Stars}}
\end{figure}

This paper presents stellar parameters and abundances for 848 giant and horizontal branch stars (Figure \ref{fig:CMD_All_Stars}). We are confident of the nitrogen abundances for stars with temperatures less than 4900~K, creating a subset of 557 stars. Section \ref{sec:sources} gives a brief description of the data sources and determination of \Teff\ and \logg. Section \ref{Sec:OFe} describes how the \OFe\ was estimated, while Section \ref{sec:method} outlines the spectral matching process briefly (see \paperone\ for more details). Section \ref{sec:Results} gives the results and compares them to previous results from \paperone\ and other researchers. Section \ref{sec:ClusteringAnalysis} shows how the stars were placed into evolutionary groups. Section \ref{sec:Discussion} discusses the abundances determined and how these fit with the different evolutionary models of \ocen.

\section{Photometric and spectroscopic data sources}\label{sec:sources}

The spectroscopic library, photometry, model atmospheres and linelists used are the same as in \paperone, where they are described in more detail. Here a brief synopsis is provided.

The \citet[][hereafter \vL]{vanLoon2007} dataset is a spectral library of 1518 post-main sequence stars in \ocen. Spectra were obtained with the 2dF instrument at the Anglo-Australian Telescope, covering approximately $\lambda\sim$3840--4940~\AA\ at a resolving power of $\lambda/\Delta\lambda\sim1600$ and with a signal-to-noise per pixel ranging from $\sim50$ in the blue to $>100$ in the redder part of the wavelength range. We use the unique ID from \citet{vanLeeuwen2000} (known as the LEID) to refer to the stars.

The \Teff, \logg\ and [Fe/H] values were used to generate the necessary model atmospheres via interpolation\footnote{From https://github.com/andycasey/atmosphy.} within the $\alpha$-enhanced grid of \citet{Castelli2003}. A value of [$\alpha$/Fe]$=0.3$ was used \citepalias{Johnson2010}. For the CH and CN region of the spectrum, line lists were from \citet{Norris2012}. These line lists were found to match to the high resolution atlas of Arcturus \citep{Hinkle2000} using abundances determined by \citet{Decin2004}. They also matched the Arcturus spectrum convolved to the resolution of the \citetalias{vanLoon2007} spectra. The atomic $\log gf$ values in the line list for the barium region investigated (400 to 500 nm) were adjusted so that the synthetic spectrum matched the Arcturus spectrum, again at both resolutions.

The \vk\ colour index was used to determine temperatures and surface gravities for the stars. This colour was selected as its colour-temperature relationship in \citet{Alonso1999,Alonso2001} does not have a strong \FeH\ dependence like \bv. For \bv, across the metallicity range of \ocen\ there could be up to a 260~K range in the \Teff\ using the \citet{Alonso1999} equations. While for \vk, the temperature range is about 20~K..

Additional photometry from the Two Micron All Sky Survey ({\sc 2mass}) \citep{Skrutskie2006} was required for the $K$ photometry of the stars. Also, high-precision CCD photometry from \citet[][hereafter \bo]{Bellini2009a} was used for their $U$, $B$ and $V$ magnitudes where a positional match between \bo\ and \vL\ existed. This photometry was used as it provided a very precise colour-magnitude diagram. The {\sc 2mass} photometry has uncertainties of about $0.2$ percent for these stars, and the \bo\ photometry has uncertainties of about $0.1$ percent. Based upon propogation of uncertainty theory, this would equate to an uncertainty of $<1$ percent in the \Teff\ for most of the stars.

Four of the stars were analyzed using their \bv\ colours: LEIDs 33062, 35250, 44262, and 44420. Their \vk\ colours were all greater than 4.7 magnitudes, which was larger than the constraints allowed for by \citet{Alonso1999}, when using an input metallicity of slightly metal-richer than $\FeH=-1.5$, which was used as a default value for all stars. These four stars were analyzed separately with their \bv\ photometry used in conjunction with a $\FeH=-1.0$, as they were found on the metal-rich red giant branch (RGB).

The positional matching between \vL, \bo\ and 2MASS was unsupervised and simply returned the closest positional match. There were some sensible magnitude cuts to remove main sequence stars. However there was no requirement that the match have similar photometry. This could mean that some stars could be matched to an incorrect star. Inspection of spectra and their matches found very few examples of stars which obviously had the wrong temperature. The indicator of this was the strength of the H$\delta$ line at 4102~\AA, which is temperature sensitive. Less than ten hot stars were found to have a hydrogen line that was much weaker than other stars of the same calculated temperature. This indicated that the temperatures determined were incorrect for these ten stars.

A catalogue of 1043 stars, with colours within the range of the \citet{Alonso1999} constraints, was created from the full library of 1518 stars of \vL. The next stage was to estimate the \OFe\ abundance of each star.

\section{\OFe\ estimation from CN and CH indices}\label{Sec:OFe}

In \paperone, the stars were selected from \vL\ spectral library so as to also have \OFe\ and \CaFe\ from \jp. However for this analysis of the complete \vL\ library, these abundances were not known for all the stars. In the case of \CaFe, a simple average value can be used. From \jp\ it was found that there was an overall mean \CaFe\ of $0.3\pm0.1$, with a very small positive correlation with metallicity. For stars with $\FeH<-1.8$, $\avg{\CaFe}=0.29$; $-1.8>\FeH>-1.6$, $\avg{\CaFe}=0.25$; $-1.6>\FeH>-1.4$, $\avg{\CaFe}=0.33$; $-1.4>\FeH>-1.2$, $\avg{\CaFe}=0.36$; $\FeH>-1.2$, $\avg{\CaFe}=0.31$. For all the stars, $\CaFe=0.3$ was therefore used. The effect of having the incorrect \CaFe\ abundance for a star is that the spectral matching pipeline will determine a metallicity that will retain the equivalent width of the Ca~I line, i.e., if \CaFe\ was incorrect by $+0.2$~dex, then the $\Delta\FeH$ found will be $-0.2$~dex.

\begin{figure}
\includegraphics[width=84mm]{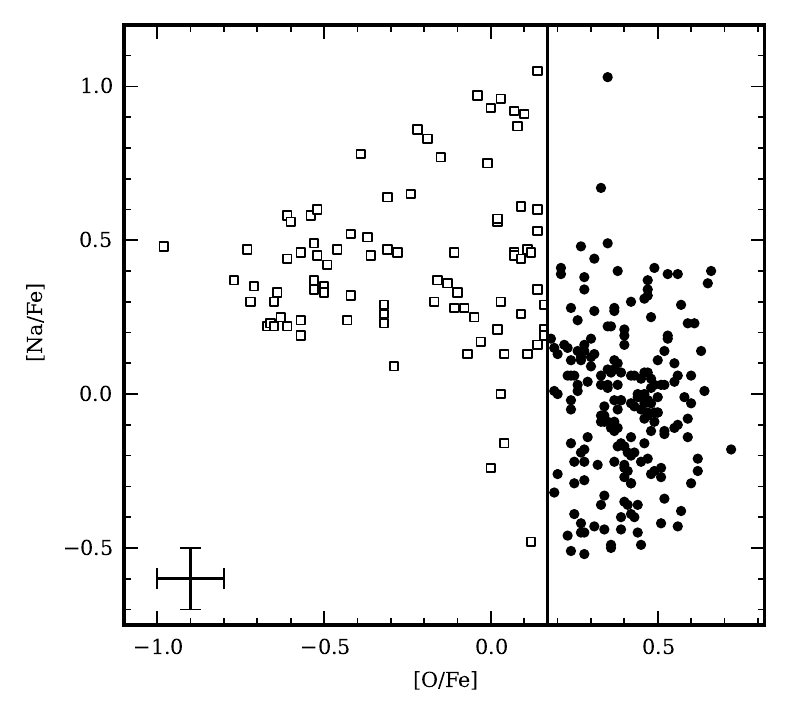}
\caption{The \OFe\ and \NaFe\ determined by \citetalias{Johnson2010} for the 291 stars in common between it and \citetalias{vanLoon2007}. The vertical line is at $\OFe=0.17$, which was used as the divider between oxygen-poor ($\square$) and oxygen-rich ($\bullet$) stars. The uncertainty is taken from  \citetalias{Johnson2010}. \label{fig:NaO}}
\includegraphics[width=84mm]{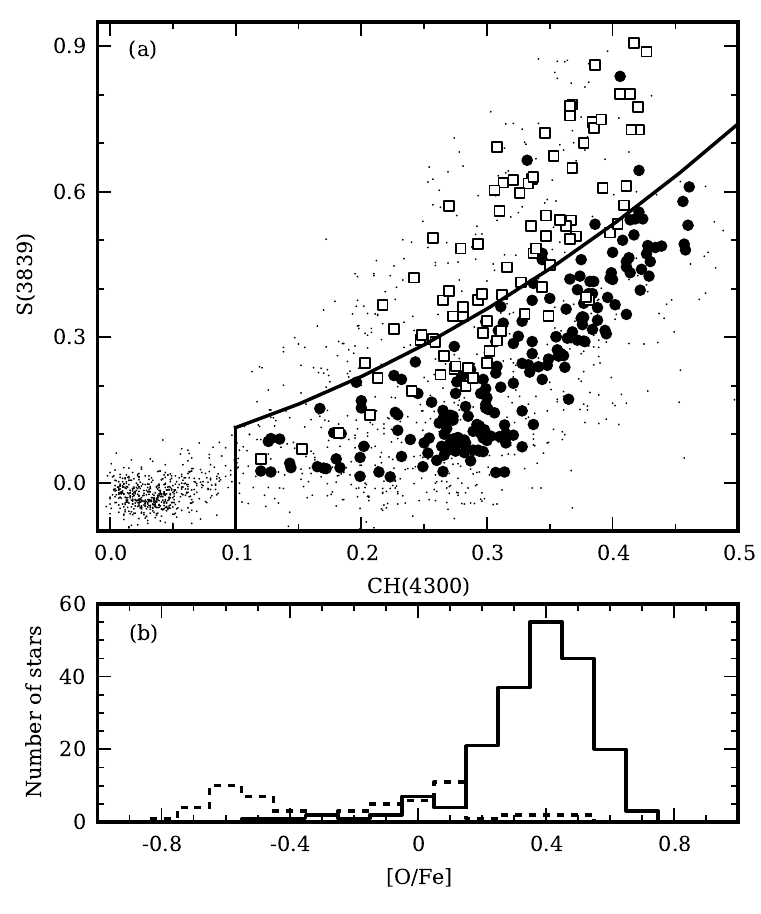}
\caption{(a) Positions of stars with $\OFe>0.17$ ($\bullet$) and $\OFe<0.17$ ($\square$) (from Figure \ref{fig:NaO}) on the plane of CH and CN indices as found by \vL. The line is a parabola that was selected to have as many of the O-rich stars below the line as possible. The small dots are the full \vL\ library. The concentration of stars with low values of both indices are the horizontal stars, which are separated by a vertical line. (b) Histogram of the \OFe\ abundance of stars above (solid line) and below (dashed) the line of (a).  \label{fig:S3839_CH4300}}
\end{figure}

The \OFe\ abundance required a more nuanced approach. In Figure \ref{fig:NaO} are the 291 stars that are in common between \vL\ and \jp. There is a 1.5~dex range of \OFe\ (\jp; \citealt{Marino2011}), and as found in \paperone, there is a degeneracy between metallicity and oxygen abundance. At a particular metallicity, there could be the maximum range of \OFe, with the intermediate metallicity stars being both the most oxygen-rich and the most oxygen-poor stars. Along with this range of metallicity, oxygen is involved in molecular equilibria with the two elements of interest: carbon and nitrogen. Unlike an atomic species where changing the abundance will most often not affect the line strength of other elements, changing the abundance of oxygen in the star results in changes in the strength of the CH and CN molecular features. For instance, using synthetic spectra, it was found that for a 3750~K star with $\FeH=-1.7$, changing the oxygen abundance by 0.5~dex required an increase in the carbon abundance of 0.4~dex to fit the CH molecular feature at $\sim4300$~\AA. The effect on the abundance results of having the incorrect \OFe\ for the star was investigated. It was found that it would not affect the \FeH\ determined, but the \CFe\ would change by an equal but opposite amount to compensate. There was no effect on the \NFe\ abundance determined.

For all the stars in the \vL\ library, there was a measurement of the S(3839) and CH(4300) indices\footnote{It should be noted that these values used \vL\ normalized spectra.}. The first measured the CN band heads that extend blueward from 3883~\AA\ and the second measured the CH band at 4300~\AA\ (the G band). (See \citet{Harbeck2003} for the definitions used by \vL.)

There exists a correlation between these two indices and \OFe. As shown in Figure \ref{fig:S3839_CH4300}a, the oxygen-rich stars are found at lower S(3839) values than the oxygen-poor stars. This comes about because the oxygen-rich stars tended to be the nitrogen-poor stars and vice versa. Using information the spectral indices of the O-rich and O-poor stars from \jp, a divider was selected that best separated the O-rich stars from the O-poor stars. It was decided to make sure that all the oxygen-rich stars were identified, at the loss of identifying some oxygen-poor stars as oxygen-rich (Figure \ref{fig:S3839_CH4300}b). This was because we wanted to include the very high S(3839) index value O-rich stars. There were 27 stars (of 99 in total: 27~per~cent) which were O-poor ($\OFe<0.17$) which were below the parabola of Figure \ref{fig:S3839_CH4300}a. They had a median \OFe\ of $-0.03$. There were 6 O-rich stars (of 222 in total: 3~per~cent) that were above the parabola, with a median \OFe\ of $+0.37$. From now on in this paper, O-rich/O-poor refers to the stars below and above the parabola in Figure \ref{fig:S3839_CH4300}a.

For O-rich stars, $\OFe=+0.4$ was used as the input oxygen abundance for the spectral matching analysis. For the oxygen-poor stars, two values were initially used: $\OFe=-0.1$ and $-0.5$. So the O-poor stars were analyzed twice.

\section{Spectral matching method}\label{sec:method}
The method for determing the abundance of the stars is the same as that used in \paperone. It is briefly described here.

Each star had a \Teff, \logg, \OFe\ estimated as described in Sections \ref{sec:sources} and \ref{Sec:OFe}. For the oxygen-rich and oxygen-poor stars there is a known range of \CFe\ and \NFe\ from \paperone. These were used to provide a random initial value for \CFe\ and \NFe, along with an \FeH\ initial value that was randomly chosen from the known range of \ocen. A model atmosphere was interpolated from a grid by \textsc{atmosphy}\footnote{https://github.com/andycasey/atmosphy} using these input values. This was then used to create a synethic spectrum using \textsc{moog}\footnote{http://www.as.utexas.edu/$\sim$chris/moog.html} \citep{Sneden1973}. The raw spectrum was continuum normalized using three anchor points: 4088~\AA, 4220~\AA\ and 4318~\AA\ \citep[as also used in][]{Worley2012}. These three points were joined by two straight lines and the intensities were mapped from the raw spectrum onto a synthetic spectrum.

The first stage of the spectral matching process was to minimize the equivalent width difference between the synthetic and observed spectra of the CaI line at 4226~\AA. This was done by changing the \FeH\ in the model until the difference could not be minimized further. The $\chi^{2}$ of the CH band head (4295--4325\AA) was minimized by changing the \CFe\ of the model. Finally the CN band head in the region 4195--4215~\AA\ was minimized. This new triplet of \FeH, \CFe\ and \NFe\ was then cycled back through to minimize the CaI feature, then the CH and then the CN. The abundances were accepted once the method converged onto the same triplet of values on two consecutive cycles.

To determine the \BaFe\ abundance, the BaII line at 4554~\AA\ was used, minimizing the difference between the observed and synethic spectrum using the \FeH\ determined for the other region of the spectrum. There was no inclusion of hyperfine splitting as it was found that at the resolution of the \vL\ spectra there was little difference in the synthetic spectra that included and excluded hfs.

There were 1043 stars for which a \Teff, \logg, \FeH, \CFe, \NFe, \BaFe\ were determined. For each star, three runs with three random starting positions were undertaken at the star's \OFe\ abundances. This gave three values for \FeH, \CFe\ and \NFe\ for each \OFe\ abundance for each star. A star had its abundance accepted if at least two of these runs resulted in the same value for all three elements. This allowed for stars where one starting position would not return a reasonable fit due to the parameter space search being trapped in local minima that were not the desired global minimum. The values reported in Tables \ref{Table:Marino2012} and \ref{Table:Results} are the mode of value returned from the three runs. Of the 1043 stars, 912 (87~per~cent) stars were accepted by this criteria. About 90~per~cent of these 912 stars that were kept had returned the same abundances on all three runs.

Breaking this down into O-rich and O-poor stars, 678 of 925 (73~per~cent) O-rich stars were kept and 234 of 271 (86~per~cent) O-poor stars were kept. This result was expected as the O-rich stars include both N-poor stars and hotter stars. Both these groups are less likely to converge. The iteration method has a limit to how small the change in the strength of the spectral features can be before it discontinues the search in that particular parameter. For instance, it will not continue to change the abundance of nitrogen in the model if the $\chi^2$ changes by less than 0.00001.

The uncertainties reported in this paper were the same as those reported in \paperone. This used the stars for which there were two spectra observed by \vL\ and found what was the average spread found: [Fe/H] was $\pm0.2$~dex, [C/Fe] was $\pm0.2$~dex, [N/Fe] was $\pm0.3$~dex and [Ba/Fe] was $\pm0.6$~dex.

To recap, the 1518 stars of the full \vL\ library were reduced to 1043 stars through the constraints of \citet{Alonso1999}, which were further reduced to 912 stars by the requirement for the spectral matching method to return the same abundances from at least two of three random starting positions in its search.

\section{Cluster membership \& quality assurance}\label{sec:Results}
Cluster membership was determined using the radial velocity \citepalias{vanLoon2007} and membership probability \citepalias{Bellini2009a}. A star was classified as a cluster member if its radial velocity was $185<V_{\rm rad}<275$ km/s ($3\sigma$); this excluded 12 stars of the 912. The membership probabilty of \bo\ was only available for those stars that had positional matches to their photometric library. We selected 90~per~cent as our cutoff for this based upon \bo's figure 12 which showed that the bulk of stars in the magnitude range had membership probabilities above 90~per~cent. These two criteria cut that number of stars to 848.

\begin{figure*}
\includegraphics[width=168mm]{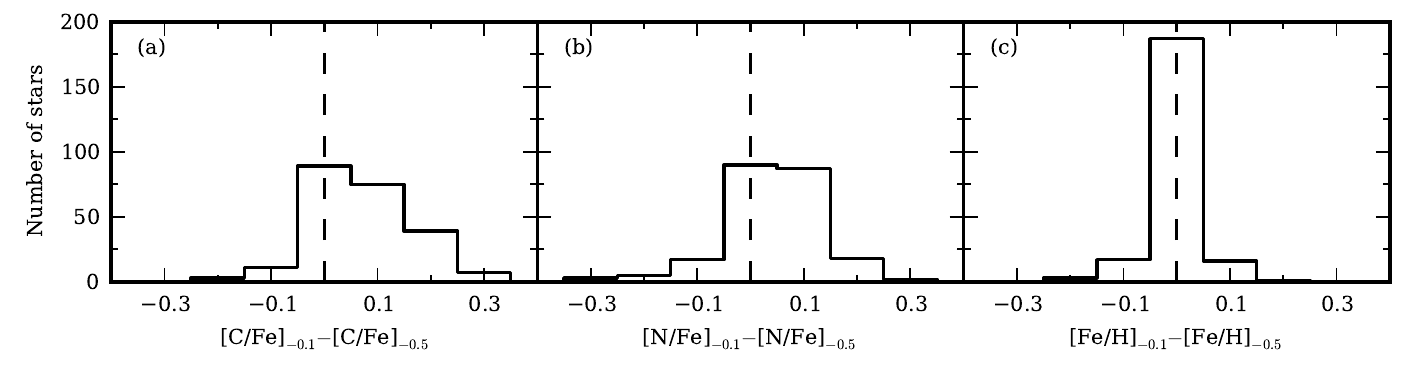}
\caption{Difference in (a) \CFe, (b) \NFe\ and (c) \FeH\ found using $\OFe=-0.1$ and $-0.5$. These show that the different \OFe\ input abundances do not significantly affect the output values of \CFe, \NFe\ and \FeH: 78~per~cent, 87~per~cent and 98~per~cent, respectively, are within 0.1 dex. This justifies the adoption of a single oxygen abundance of $\OFe=-0.5$ in the rest of this paper for the O-poor stars. \label{fig:CFe_OCompare}}
\includegraphics[width=168mm]{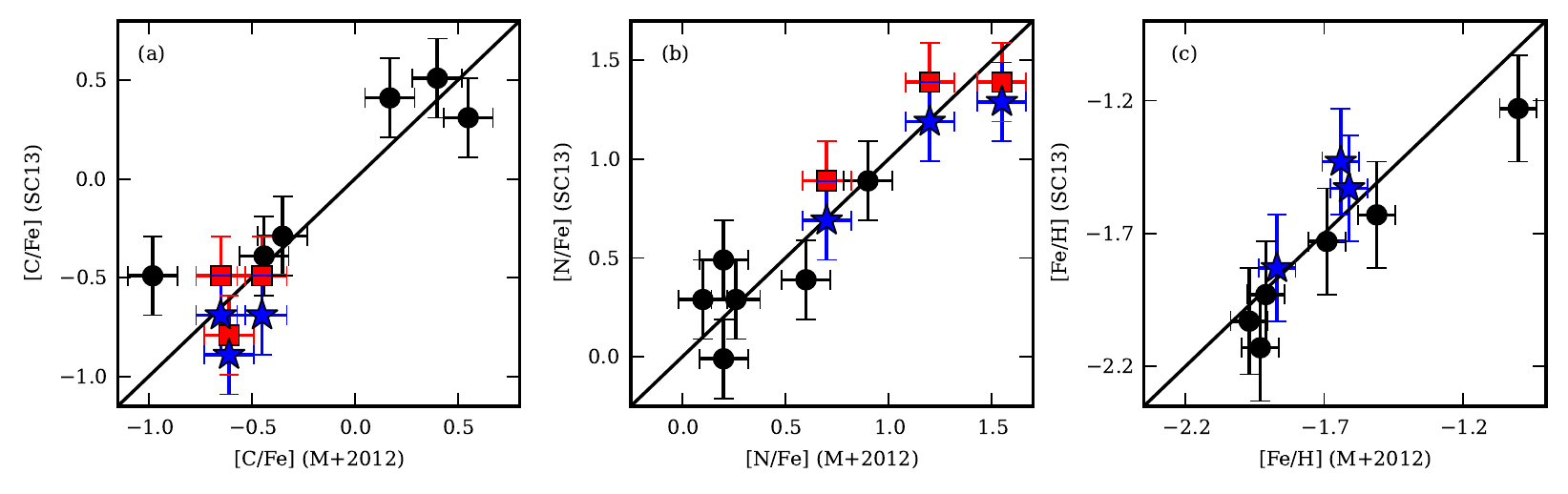}
\caption{The nine stars in common between this work and \citetalias{Marino2012} were used to determine systematic offsets. (a) \CFe\ values were adjusted by $+0.41$~dex, (b) \NFe\ by $-0.51$ and (c) \FeH\ by $+0.09$. In (a) and (b), the abundances found for the O-poor stars using $\OFe=-0.1$ (red circles) and $-0.5$ (blue stars) are shown. Those at the same [X/Fe] from \citetalias{Marino2012} are the same star. For \FeH, both oxygen-poor abundances returned the same \FeH\ for these particlar stars. \label{fig:M2012_FeH}}
\end{figure*}

For the oxygen-poor stars, there were two values of \FeH, \CFe\ and \NFe\ due to the use of two \OFe\ abundances. It did not make sense to simply take the average. The reason for using two values was that it was unknown how oxygen poor the star was using its CH(4300) and S(3839) indices. Taking the average of the \FeH, \CFe\ and \NFe\ abundances would not return the correct value for those stars that were at the extremes of the oxygen abundance. Inspection of the histograms in Figure \ref{fig:CFe_OCompare} showed that there was not a large difference between the \CFe, \NFe, \FeH\ determined with $\OFe=-0.1$ and $-0.5$. For \FeH, 83~per~cent of the stars returned the same metallicity value and 98~per~cent were within $\pm0.1$ dex. For \CFe, 78~per~cent were within $\pm0.1$ dex and for \NFe\ it was 87~per~cent. For this reason it was decided to use only one oxygen abundance ($\OFe=-0.5$) for the oxygen-poor stars in the subsequent analysis presented in this work. As stated previously, for the oxygen-rich stars, $\OFe=+0.4$.

\begin{table*}
\begin{minipage}{135mm}
 \caption{Stellar parameters determined for the nine stars in common between \citetalias{Marino2012} and this work ($\equiv$~SC13). The LEID is the star label given by \citetalias{vanLoon2007} and the ID is the numbering scheme from \citetalias{Marino2012}. The [Fe/H], [C/Fe], [N/Fe] from \citetalias{Marino2012} are shown for comparison.}
 \label{Table:Marino2012}
 \begin{tabular}{rcrrrrrrrr}
 \hline
 LEID & ID & \multicolumn{2}{c}{[Fe/H]} & \multicolumn{2}{c}{[C/Fe]} & \multicolumn{2}{c}{[N/Fe]} & \multicolumn{2}{c}{[O/Fe]} \\
 \citetalias{vanLoon2007}     &  \citetalias{Marino2012} & \citetalias{Marino2012} & SC13 & \citetalias{Marino2012} & SC13 & \citetalias{Marino2012} & SC13 & \citetalias{Marino2012} & SC13  \\
 \hline
33139 & 250606 & $-1.00$ & $-1.3$ &    0.55 & $-0.1$ & 0.60 & 0.9 &    0.86 & $+0.4$\\
35090 & 246585 & $-1.64$ & $-1.5$ & $-0.65$ & $-1.1$ & 1.55 & 1.8 & $-0.25$ & $-0.5$\\
36156 & 245724 & $-1.97$ & $-2.1$ & $-0.44$ & $-0.8$ & 0.20 & 0.5 &    0.30 & $+0.4$\\
37253 & 242745 & $-1.91$ & $-2.0$ & $-0.35$ & $-0.7$ & 0.10 & 0.8 &    0.37 & $+0.4$\\
38115 & 241359 & $-1.69$ & $-1.8$ &    0.17 &    0.0 & 0.26 & 0.8 &    0.40 & $+0.4$\\
46150 & 224500 & $-1.51$ & $-1.7$ &    0.40 &    0.1 & 0.20 & 1.0 &    0.49 & $+0.4$\\
46194 & 225246 & $-1.87$ & $-1.9$ & $-0.61$ & $-1.3$ & 0.70 & 1.2 &    0.19 & $-0.5$\\
48235 & 220325 & $-1.61$ & $-1.6$ & $-0.45$ & $-1.1$ & 1.20 & 1.7 &    0.00 & $-0.5$\\
51091 & 215367 & $-1.93$ & $-2.2$ & $-0.98$ & $-0.9$ & 0.90 & 1.4 &    0.15 & $+0.4$\\
\hline
 \end{tabular}
\end{minipage}
\end{table*} 

Of the 77 stars in \citetalias{Marino2012}, nine\footnote{LEID36179 (ID 244812; \citealt{Marino2011}) also matched but was excluded due to the extreme mismatch in its \NFe, as also found in \paperone.} were in common with the 848 stars presented in this research (Figure \ref{fig:M2012_FeH} and Table \ref{Table:Marino2012}). These nine stars provide a direct comparison for Fe, C and N, with the caveat that the abundances were determined with different \OFe\ in each study. Assuming a straight line with a gradient of unity, the best fitting intercept for \FeH\ is $+0.09$, \CFe\ is $+0.47$ and \NFe\ is $-0.51$. All subsequent values of \FeH, \CFe\ and \NFe\ have had these offsets applied (with the values rounded to one decimal place).

\begin{figure}
\includegraphics[width=84mm]{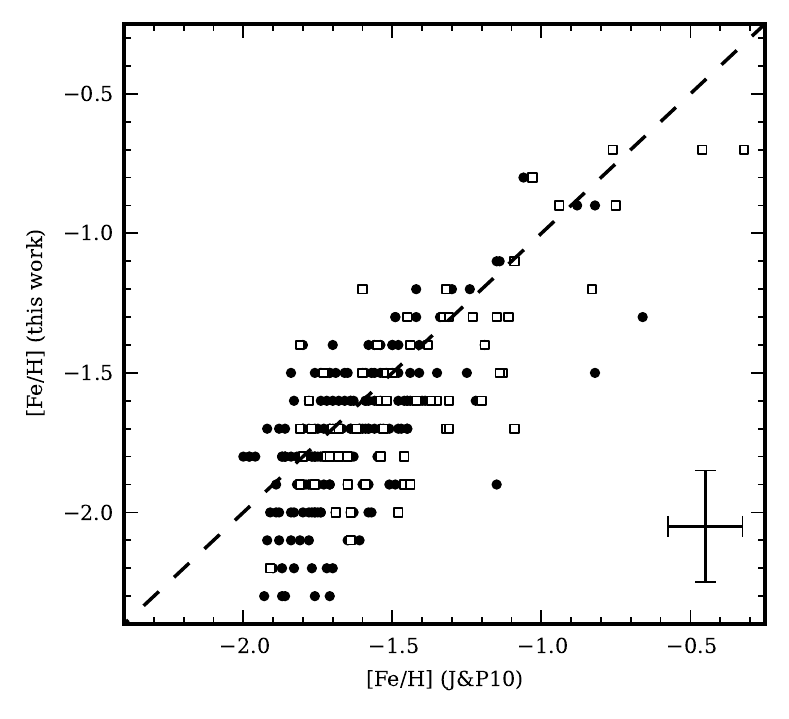}
\caption{Comparison of the \FeH\ values found for the same stars by this work and those of \citetalias{Johnson2010}. The solid circles are the O-rich stars and the squares are the O-poor stars. The dashed line is the one-to-one line. \label{fig:FeH_JP10Compare}}
\includegraphics[width=84mm]{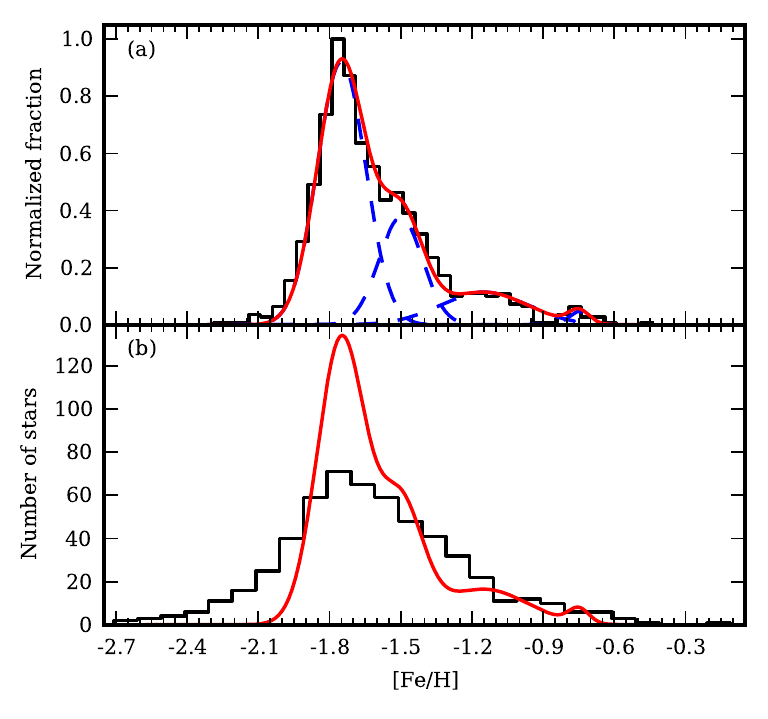}
\caption{(a) Metallicity histogram from \citetalias{Johnson2010}, with a solid red line showing the best fitting result of adding together four gaussians with $\mu$ of $\FeH=-1.75$, $-1.50$, $-1.15$, $-0.75$ (dashed blue curves). (b) \FeH\ histogram of this study with the solid line from the top panel for guidance.\label{fig:FeH_Histo}}
\end{figure}

In Figures \ref{fig:FeH_JP10Compare} and \ref{fig:FeH_Histo}, the \FeH\ of this work and \jp\ are shown and compared. There is some degeneracy in our values, but this is expected due to the lower resolution. We also find that the metallicity distribution is similar but broadened due to the lower resolution. The strength of the peak at $\FeH=-1.75$ is not as great in our data, with a fraction of stars found at lower and higher metallicities that should be found in this main metallicity component of the cluster.

\begin{table*}
\begin{minipage}{152mm}
\caption{Stellar parameters determined for the 912 stars. For each star an identifier from \vL\ (LEID) and 2MASS is given. For stars that also matched to \bo, their identifier is provided. In the full version of the table, the non-cluster members described in Section \ref{sec:ClusteringAnalysis} are appended.}
\label{Table:Results}
\begin{tabular}{cccccrrrrrc}
\hline
LEID & 2MASS & \bo & \Teff (K) & \logg & \FeH & \CFe & \NFe & \BaFe & \OFe & Group\\
\hline
42044 & 13260536-4728206 & 200024 & 3650 & 0.3 & $-1.5$ & 0.2 & 2.5 & 0.7 & 0.4 & 4\\
35094 & 13262881-4725235 & 259071 & 3700 & 0.7 & $-1.2$ & $-0.4$ & 2.6 & 1.3 & 0.4 & 4\\
48150 & 13263981-4731069 &  & 3750 & 0.3 & $-1.9$ & 0.9 & 1.2 & 1.5 & 0.4 & 2\\
25065 & 13270118-4720409 &  & 3850 & 0.6 & $-1.3$ & $-0.9$ & 2.4 & 0.3 & $-0.5$ & 4\\
48120 & 13263304-4731003 &  & 3850 & 0.3 & $-1.7$ & 0.4 & 1.2 & 0.2 & 0.4 & 2\\
47399 & 13271864-4730509 & 149466 & 3900 & 0.5 & $-1.6$ & 0.1 & 0.8 & 0.1 & 0.4 & 2\\
\hline
\end{tabular}
(This table is available in its entirety in a machine-readable form in the online journal. A portion is shown here for guidance regarding
its form and content.)
\end{minipage}
\end{table*}

Inspection of the full set of spectral fits showed that there was a temperature limit, above which there was little sensitivity to changes in the iron, carbon and nitrogen abundances in the model atmopshere. A value of $\Teff\leq4900$K was selected as the temperature limit. A further limit was imposed that CH(4300) must be greater than 0.1 to exclude any horizontal branch stars (see Figure \ref{fig:S3839_CH4300}a). This created a subset of 557 stars (of the 848 cluster members, the subset of the 1015 stars which had abundannces determined and also had their three spectral matching runs converge), which are the stars that will be plotted on subsequent figures. These are the stars which had abundances determined from the input catalogue of 1015 stars with photometry and within the colour range of \citet{Alonso1999}. Table \ref{Table:Results} presents the abundance results for the 848 stars.

\section{Clustering analysis}\label{sec:ClusteringAnalysis}

\begin{figure}
\includegraphics[width=84mm]{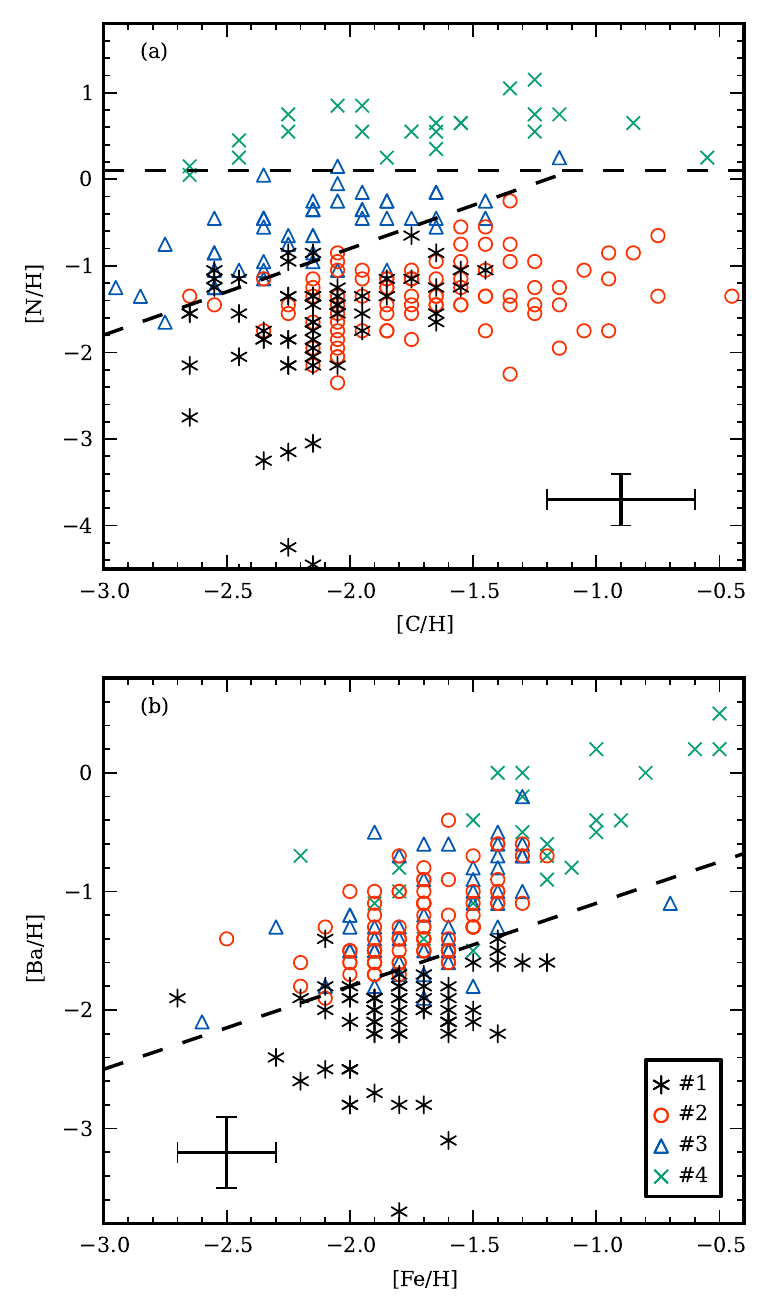}
\caption{(a) \NH\ versus \CH\ and (b) \FeH\ versus \BaH\ from \paperone, with the stars grouped as found in that paper using k-means clustering analysis. These planes were used to define the groups in this current paper, with the dashed lines showing the divisions between each group. In (a), above the horizontal line is group \#4, between that line and the diagonal line is group \#3. All the stars below this are groups \#1 and \#2 and they are divided in (b) by the diagonal line (above is group \#2 and below is group \#1.  \label{fig:GroupsHydrogen}}
\end{figure}

From \paperone\ it was known that there were at least four distinct groups of stars based upon their abundances. Each group had distinguishing abundance characteristics (Table \ref{Table:Qualitative}) that could be used to place other stars into groups without rerunning the k-means clustering analysis that was used in \paperone. For instance, the stars of group \#4 were all much more nitrogen rich than any other stars of the cluster. Group \#3 could be distinguished from groups \#1 and \#2 by its larger nitrogen abundance. In the carbon-nitrogen plane, groups \#1 and \#2 could not be distinguished easily but are separate in the [Ba/H]-\FeH\ plane (Figure \ref{fig:GroupsHydrogen}).

It was decided that instead of using the k-means analysis on this new sample of stars, that criteria would be developed to place stars into groups. This had the advantage of using abundances that would be well-defined for some stars, but not well-defined for others. For example the low-N stars, for which \NFe\ was not very precise, were distinguished using barium abundances.

Group \#4 stars were those for which $\NH>0.1$, group \#3 stars were those remaining stars where $\NH>\CH+1.2$, group \#2 were those remaining stars where $\BaH>0.7\times\FeH-0.4$, and group \#1 were the rest. These divisions can be seen in Figure \ref{fig:GroupsHydrogen}.

Using these criteria on the results from \paperone\ resulted in essentially the same group classifications. Of the new group \#1 stars, four were old group \#2 and the rest were group \#1. In the new group \#2, all but five were originally group \#2 and the rest were group \#1. The new group \#3 had 2 group \#4 stars, 10 group \#1 and 6 group \#2. This still meant that 87~per~cent of stars of the new group \#3 were group \#3 in \paperone. In the new group \#4, 2 were old group \#3 stars.

\begin{table}
\caption{The cluster centres found using the grouping criteria described in Section \ref{sec:ClusteringAnalysis} for the 557 stars. For each parameter, the average and standard deviation are given, along with the number of stars in each group. The symbol is the identifier used for that group in the figures.}
\label{Table:Simpson2013Stats}
 \begin{tabular}{rcrrrrrrr}
 \hline
 \#  & Symbol  & No.  & [Fe/H] & [C/Fe] & [N/Fe] & [Ba/Fe] \\
 \hline
 1 & \textasteriskcentered                    &  79 & $-1.7$ & $-0.2$ & $0.1$ & $-0.4$\\
   &                                          &     & $\pm0.3$ & $\pm0.2$ & $\pm0.5$ & $\pm0.4$\\
 2 & \textcolor{squares}{$\circ$}           & 254 & $-1.7$ & $0.1$ & $0.3$ & $0.7$\\
    &                                          &     & $\pm0.4$ & $\pm0.4$ & $\pm0.5$ & $\pm0.4$\\
 3 & \textcolor{triangles}{$\triangle$}  & 165 & $-1.7$ & $-0.5$ & $1.1$ & $0.4$\\
    &                                          &     & $\pm0.3$ & $\pm0.2$ & $\pm0.3$ & $\pm0.6$\\
 4 & \textcolor{crosses}{$\times$}            &  59 & $-1.1$ & $-0.5$ & $1.8$ & $0.7$\\
    &                                          &     & $\pm0.4$ & $\pm0.4$ & $\pm0.5$ & $\pm0.6$\\

\hline
 \end{tabular}
\end{table}

The mean abundances and standard deviations of the grouping of the 557 stars are shown in Table \ref{Table:Simpson2013Stats}, along with the symbols and colours used for these groups on subsequent figures.

\section{Discussion}\label{sec:Discussion}
We have compiled an extensive set of atmospheres, parameters and abundances for 557 giant branch stars in \ocen. We discuss these in the context of the methods used but more importantly in terms of cluster evolutionary models.

\begin{figure}
\includegraphics[width=84mm]{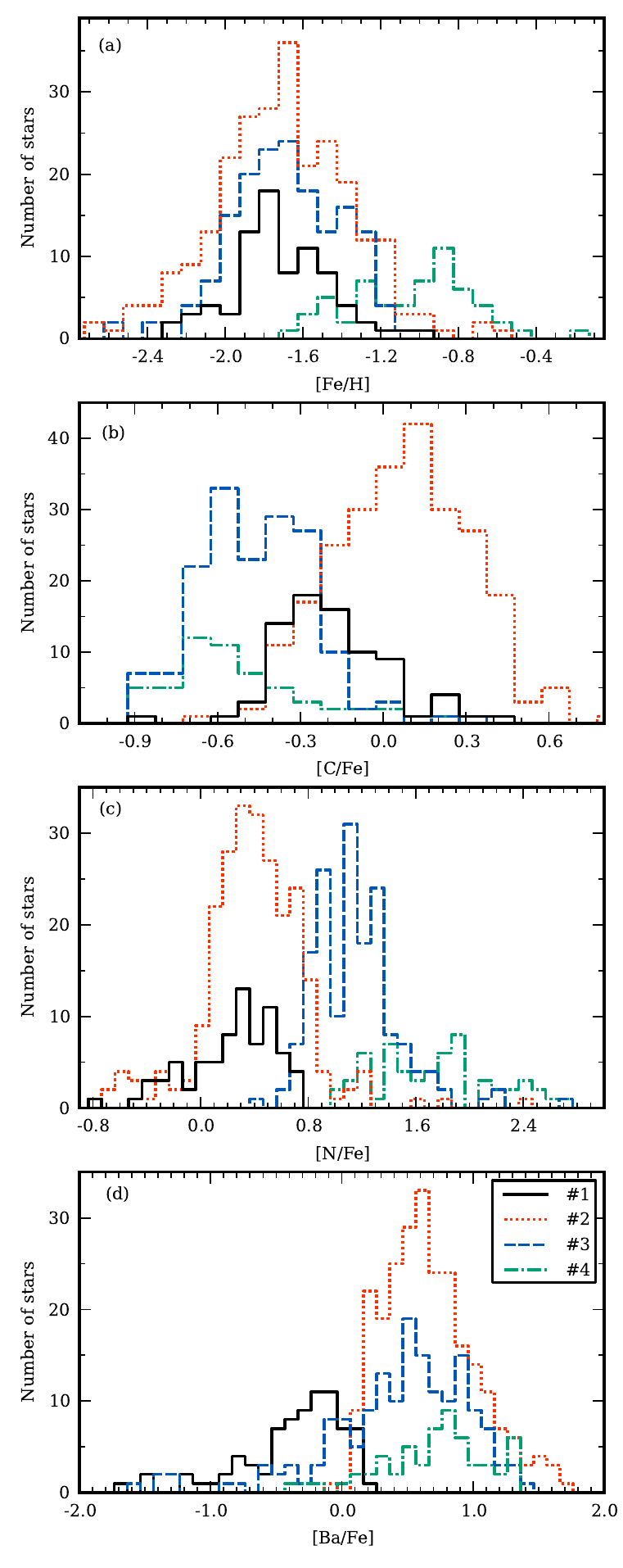}
\caption{For each of the inferred abundances, (a) \FeH, (b) \CFe, (c) \NFe\ and (d) \BaFe, histograms illustrate the concentration of stars at particular abundances. \label{fig:Abund_Histo}}
\end{figure}

The spectral resolution required the abundances inferred to be limited to 0.1 dex step sizes. This means that there are datum points that actually have several stars  at the same value. For example, at $\CFe=0.0$, $\NFe=0.4$, there are 10 stars. In order to illustrate the density of stars at a given value, we show histograms  in Figure \ref{fig:Abund_Histo}. In the case of \FeH\ (Figure \ref{fig:Abund_Histo}a), they show that group \#4 is definitely metal richer than the other groups, with hints that \#1 is slightly metal poorer than the other two groups. Conversely Figure \ref{fig:Abund_Histo}d shows that group \#1 is barium poorer than the other three groups.

\begin{figure}
\includegraphics[width=84mm]{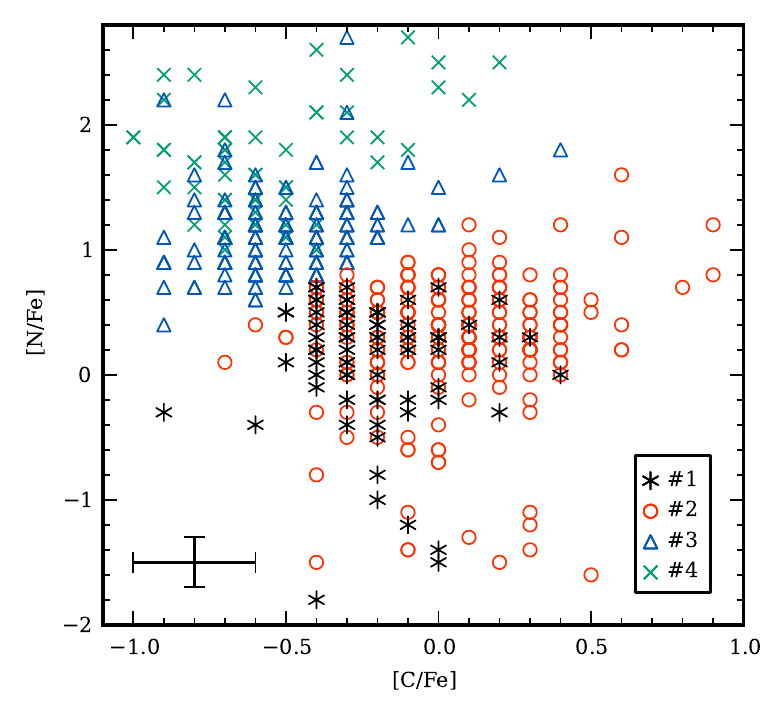}
\caption{\NFe\ and \CFe\ abundances for the four groups described in Section \ref{sec:ClusteringAnalysis} and Table \ref{Table:Simpson2013Stats}. The accompanying histograms are Figures \ref{fig:Abund_Histo}c (\NFe) and \ref{fig:Abund_Histo}b (\CFe). It shows that there are four distinct regions when investigating carbon and nitrogen: a carbon-poor and nitrogen-poor group (\#1) which is also iron-poor; a carbon-rich but nitrogen-poor group (\#2); a carbon-poor but nitrogen-rich group (\#3); and a very nitrogen-rich group (\#4). \label{fig:GroupsCNAll}}
\includegraphics[width=84mm]{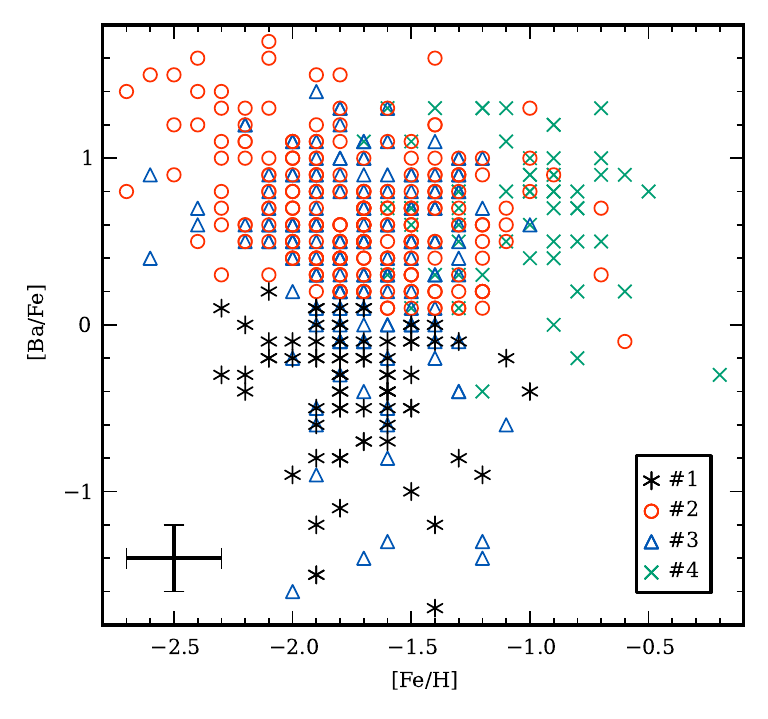}
\caption{\BaFe\ versus \FeH\ for the four groups described in Section \ref{sec:ClusteringAnalysis} and Table \ref{Table:Simpson2013Stats}. The accompanying histograms are Figures \ref{fig:Abund_Histo}d (\BaFe) and \ref{fig:Abund_Histo}a (\FeH). There are three regions in this abundance plane: a Fe-poor and Ba-poor region (group \#1); a Fe-poor to Fe-intermediate and Ba-rich region (groups \#2 and \#3); and a Fe-rich and Ba-rich region (group \#4). \label{fig:GroupsFeBaAll}}
\end{figure}

The key driver behind this research is the determination of carbon and nitrogen abundances for stars in the globular cluster \ocen\ (Figures \ref{fig:Abund_Histo}b, \ref{fig:Abund_Histo}c and \ref{fig:GroupsCNAll}). In \paperone\ it was shown that using C and N, groupings that were not previously seen, could be created. There are four broad groupings of stars: two low-carbon groups, one with higher nitrogen than the other; and two carbon-deficient but nitrogen-rich groups. It is not possible to seperate the stars of \citetalias{Johnson2010} into groups such as ours as there is no other abundance planes that provided all the information required to create the groupings that are found in carbon and nitrogen for the carbon-deficient groups, \#1 and \#2. The problem is that they have the same ranges og oxygen and sodium abundances. Compared to \paperone, we find that group \#3 extends to much carbon-poorer abundances, hinting at processing of carbon in group \#1 or \#2 (depending on the evolutionary model adopted) into nitrogen or sodium. By definition, \#4 is nitrogen rich, and has a large range of carbon abundances. Group \#4 is simply defined by its [N/H] abundance, but this selects the metal- and sodium-rich stars, which form a very separate giant branch on the colour-magnitude diagram.

The \BaFe\ results (Figures \ref{fig:Abund_Histo}a, \ref{fig:Abund_Histo}d and \ref{fig:GroupsFeBaAll}) are consistent with previously found results that the barium abundance of stars rises rapidly at low metallicities and then flattens out at higher metallicities to a constant value. The same features have been observed by \jp\ and \citet{Marino2011}. \citet{Stanford2010} found an increase with metallicity but their most metal-rich stars where at intermediate barium abundances. \citet{Villanova2010} identified a bifurcation that is not present in any other work. Their figure 4 shows a similar distribution of stars to our Figure \ref{fig:GroupsFeBaAll}. In figure 2a of \citet{Marino2011}, the distribution of barium with iron is the clearest. From the lowest metallicities to $\FeH\sim-1.5$ there is a step increase in \BaFe\ from less than $-0.5$~dex to $+0.5$~dex. The barium abundance is then essentially constant at higher metallicities. 

\begin{figure}
\includegraphics[width=84mm]{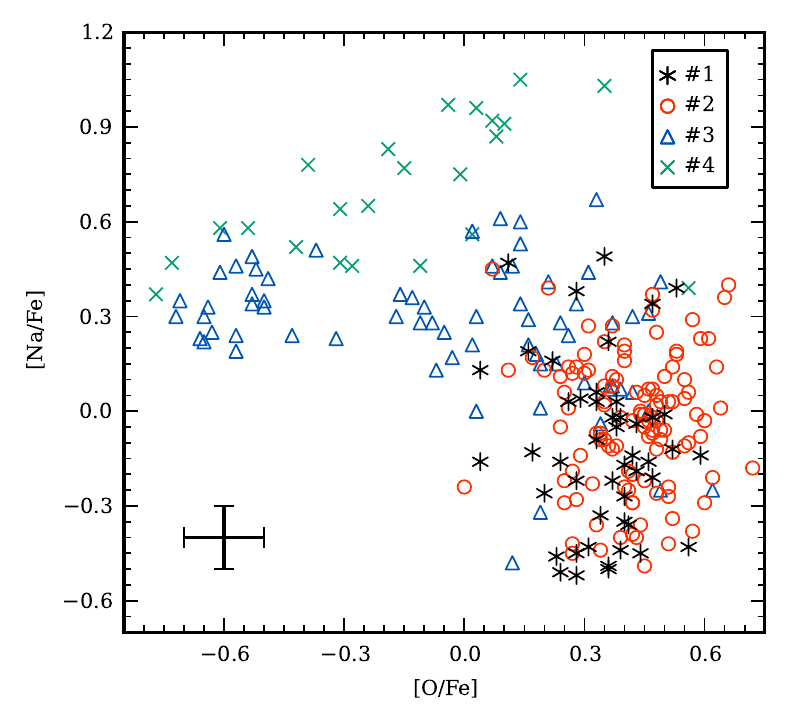}
\caption{There were 260 stars in common between this work and \citetalias{Johnson2010} and these stars are shown with their \OFe\ and \NaFe\ abundances from \jp. It was found that like \paperone, the O-rich stars consist of two populations with differing carbon and barium abundances. The Na-rich stars form a distinct population of stars in the N-C (Figure \ref{fig:GroupsCNAll}) and Ba-Fe (Figure \ref{fig:GroupsFeBaAll}) planes. \label{fig:GroupsONaAll}}
\end{figure}

The lack of \OFe\ was the driver behind the use of the CH(4300) and S(3839) indices as the estimator of oxygen abundance of the stars. We still do have \OFe\ for 261 of our 557 stars\footnote{This is more than the 221 stars reported in \paperone\ because in that work we concentrated on those stars that also had \bo's photometry.}. The resulting groupings in the O-Na plane (Figure \ref{fig:GroupsONaAll}) mimic those found in \paperone\ with the oxygen-rich stars consisting of two well-mixed groups in terms of their \OFe\ and \NaFe. The Na-rich stars of group \#4 form an sodium-oxygen correlation \citep{Johnson2010,Marino2011} that is not observed in other clusters. The sodium-oxygen anticorrelation is observed in Galactic globular clusters \citep{Carretta2009}, but the metallicity distribution of \ocen\ seems to require a modified version of this process due to the different yields that come from AGB stars of different metallicities. How the helium affects these yields also needs to be understood.

The fact that the metal-poor group \#1 stars have a lower barium abundance than the other groups implies that there is an age difference between the metal-poor group and the other groups. This is because the site of s-process production is thought to be in 3--8 $M_{\sun}$ AGB stars which have a minimum lifetime of about $10^{8}$ years. \jp\ found that there is no trend of europium with metallicity which rules out any r-process component in the production of the barium. 

\begin{figure}
\includegraphics[width=84mm]{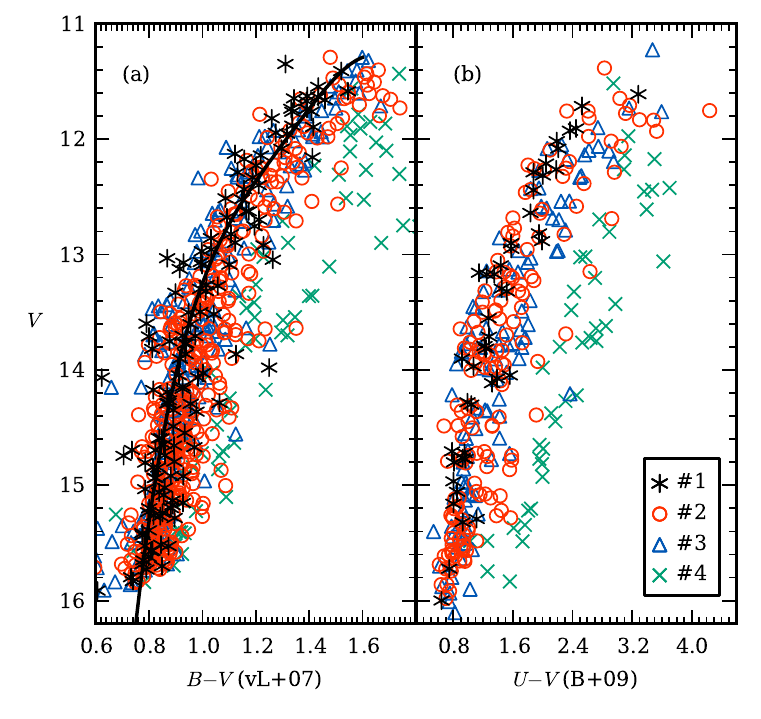}
\caption{(a) The four groups shown on a CMD using the photometry of \citetalias{vanLoon2007}. The straight line is a four-degree polynomial that is used to define a locus of the GB. (b) The four groups shown on a CMD using the photometry of \bo. There are fewer stars than in (a) because not all of the stars in \vL\ were positionally matched to \bo. \label{fig:GroupsCMDAll}}
\includegraphics[width=84mm]{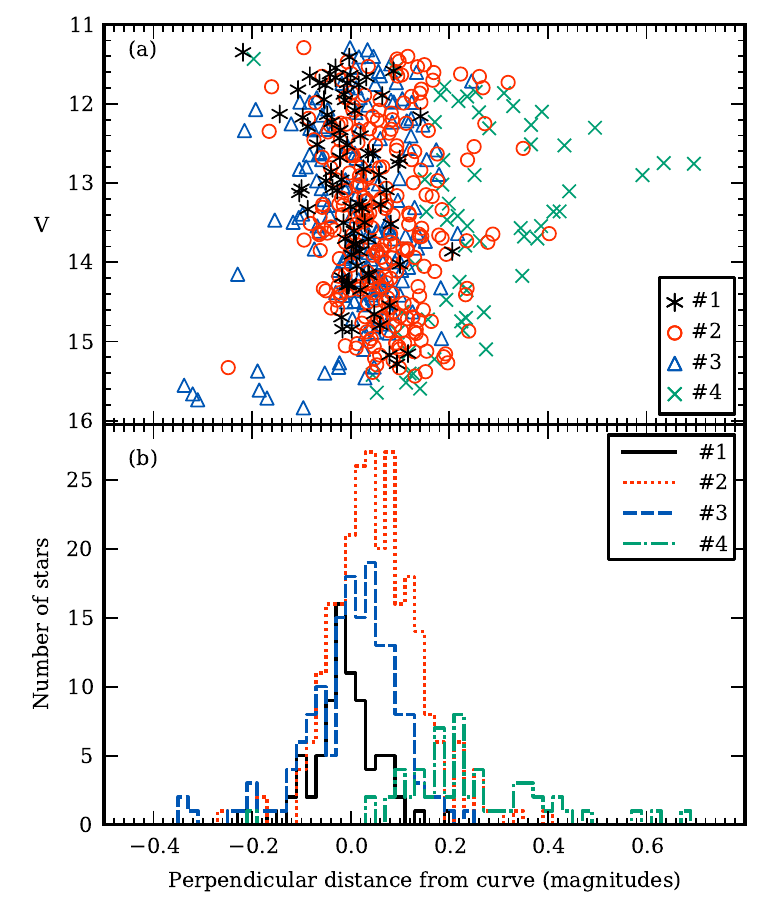}
\caption{(a) The GB locus of Figure \ref{fig:GroupsCMDAll}a was used to find the perpendicular distance of each star and then a `straightened CMD' was produced. (b) Histogram of the distances of stars from the GB locus. Group \#4 is found to be redder than the other groups, while group \#1 is on average slightly bluer than the other groups. \label{fig:GroupsCMDStraightened}}
\end{figure}

The CMD (Figure \ref{fig:GroupsCMDAll}) confirms that the nitrogen-rich group \#4 is the metal-rich RGB-a (\citealt{Pancino2000} and figure 8 of \citealt{Marino2011}). The three other groups are mixed on the giant branch. The relative positions of the groups on the CMD suggested that group \#1 was slightly bluer than the other groups. In order to investigate this further, a four degree polynomial was fitted along the edge of the giant branch. For each star, the normal distance from this line was found and this is plotted in the Figure \ref{fig:GroupsCMDStraightened}a. This normal distance was used instead of the \bv\ difference as it did not exaggerate the effects at the bright end of the GB where it flattens out to a roughly constant $V$. In Figure \ref{fig:GroupsCMDStraightened}b we present a histogram of this distance from the blue edge of the GB. It shows that group \#4 stars are definitely found at larger positive distances than the other groups. Group \#1 has an average distance of $0.00\pm0.06$ magnitudes, \#2 is $0.06\pm0.09$, \#3 is $0.00\pm0.17$ and \#4 is $0.25\pm0.15$. So there is a suggestion that group \#1 is bluer than the other groups but they are still within the uncertainty range of each other. 

In \citet{Herwig2012}, they identified galactic plane passages as a mechanism for stripping the cluster of gas. Their evolutionary sequence would have our group \#1 as the primordial population, followed by group \#3 stars forming and then group \#4 stars. The last population formed would be group \#2 from the gas of stellar winds of the group \#1 stars and after all residual gas had been stripped by a galactic plane passage. This theory has the problem that it would seem to require all globulars to be undergoing galactic plane passages in order to account for the sodium-oxygen anticorrelations. 

The same order of group formation would result from the formation model of \citet{Valcarce2011}. Group \#1 would form and pollute the environment with sodium-rich, oxygen-poor material from the winds of massive stars. The core-collapse supernovae of group \#1 would compress this gas, starting the formation of the second generation, composed of our group \#3 stars. Gas would once again accumulate at the core from the ejecta of SN of group \#1, and winds of lower-mass group \#1 and massive group \#3 stars. A third period of star formation would take place to create the group \#4 stars. Finally more gas would accumulate from the winds of all generations of stars and form the oxygen-rich, sodium-poor group \#2. One aspect they do not take into account is how the neutron-capture element profile is created. Why would the final generation of stars formed from the winds of all the previous generation of stars have the same s-process element range as the second generation of stars? It would also require the processing of nitrogen to carbon, since the group \#2 stars are much more carbon-rich than any of the other groups and also at the same carbon abundance as the group \#1 stars which form the first generation.

The self-enrichment models are the generally accepted method for creating the sodium-oxygen anticorrelations in GCs. This uses a combination of AGB star winds and supernovae to pollute the intra-cluster medium and also to expel extra gas. In \citet{DAntona2011} their model has the group \#4 forming last in the cluster, with the group \#3 forming before group \#2.

\section{Conclusions}
Using the \vL\ spectral library, stellar parameters and elemental abundances (\FeH, \OFe, \CFe, \NFe\ and \BaFe) were determined for 848 members of the globular cluster \ocen. In addition over 100 non-cluster members were also analyzed. This is one of the largest abundance analysis of this globular cluster in the literature, complementing the work of \jp. With limits imposed due to the spectral resolution of the spectra, we are confident of the nitrogen abundances for 557 stars with $\Teff\leq4900$ K.

This work confirmed what was found in \paperone\ regarding the relative abundances of four different groups of stars. We identified a metal-poor group of stars (our group \#1) that is oxygen rich, while poor in sodium, nitrogen and carbon. This group exists on the bluest edge of the giant branch of the colour-magnitude diagram, which matches to its metallicity. We believe this to be the oldest extant population in the cluster.

Groups \#2 and \#3 form the bulk of the stars in our sample and probably the cluster. In \paperone\ we showed that these stars have similar ranges of barium and iron, potentially making them coeval or one forming from the ejecta of the other group. This requires processes that change the abundance of oxygen, sodium, carbon and nitrogen, while not affecting the metallicity or s-process abundance of the stars. These two groups form the extended Na-O anticorrelation that is typical of a globular cluster. Of course these groups show a metallicity range that is not present in a simple-stellar population globular cluster.

Group \#4 are the extreme stars of our sample. They have the highest Fe, N, Na, Ba of all the stars. They are potentially the last population of stars to form in the cluster, due to their high iron content. Even if not the last population of stars to form, they are certainly one of the later groups. Their Na-O correlation \citep{Johnson2010,Marino2011} is not observed in other globular clusters. This could require the mass of \ocen\ to retain the gas that would normally be lost in other clusters but here was kept and could form this generation of stars.

Due to the biased nature of our sample of stars, we have avoided drawing conclusions about the radial aspects of these groups. We find evidence that the nitrogen-rich stars are not found at large radial distances in the cluster, being more centrally concentrated. This matches to the work of \jp\ who found that O-poor stars are similarly centrally concentrated. This would make them our groups \#3 and \#4 stars, which have both low oxygen and high nitrogen. Which is the driver here? Are they nitrogen rich because they are oxygen poor, or vice versa?

\section*{Acknowledgments}
The authors were supported by the Marsden Fund Council from New Zealand Government funding, administered by the Royal Society of New Zealand. JDS was also funded by the University of Canterbury. We also thank Andrew Ridden-Harper for the CN and CH matching code; Andy Casey for a Python wrapper for MOOG; Clare Worley for her input into the final version of this manuscript.

This publication makes use of data products from the Two Micron All Sky Survey, which is a joint project of the University of Massachusetts and the Infrared Processing and Analysis Center/California Institute of Technology, funded by the National Aeronautics and Space Administration and the National Science Foundation.

\bibliographystyle{mn2e}
\bibliography{references}

\begin{thebibliography}{}

\bibitem[\protect\citeauthoryear{{Alonso}, {Arribas} \&
  {Mart{\'i}nez-Roger}}{{Alonso} et~al.}{1999}]{Alonso1999}
{Alonso} A.,  {Arribas} S.,    {Mart{\'i}nez-Roger} C.,  1999, \aaps, 140, 261

\bibitem[\protect\citeauthoryear{{Alonso}, {Arribas} \&
  {Mart{\'i}nez-Roger}}{{Alonso} et~al.}{2001}]{Alonso2001}
{Alonso} A.,  {Arribas} S.,    {Mart{\'i}nez-Roger} C.,  2001, \aap, 376, 1039

\bibitem[\protect\citeauthoryear{{Anderson}}{{Anderson}}{1997}]{Anderson1997}
{Anderson} J.,  1997, PhD thesis, University of California, Berkeley

\bibitem[\protect\citeauthoryear{{Bedin}, {Piotto}, {Anderson}, {Cassisi},
  {King}, {Momany} \& {Carraro}}{{Bedin} et~al.}{2004}]{Bedin2004}
{Bedin} L.~R.,  {Piotto} G.,  {Anderson} J.,  {Cassisi} S.,  {King} I.~R.,
  {Momany} Y.,    {Carraro} G.,  2004, \apjl, 605, L125

\bibitem[\protect\citeauthoryear{{Bellini}, {Piotto}, {Bedin}, {King},
  {Anderson}, {Milone} \& {Momany}}{{Bellini} et~al.}{2009}]{Bellini2009a}
{Bellini} A.,  {Piotto} G.,  {Bedin} L.~R.,  {King} I.~R.,  {Anderson} J.,
  {Milone} A.~P.,    {Momany} Y.,  2009, \aap, 507, 1393

\bibitem[\protect\citeauthoryear{{Cannon} \& {Stobie}}{{Cannon} \&
  {Stobie}}{1973}]{Cannon1973}
{Cannon} R.~D.,  {Stobie} R.~S.,  1973, \mnras, 162, 207

\bibitem[\protect\citeauthoryear{{Carretta}, {Bragaglia}, {Gratton},
  {Lucatello}, {Bellazzini}, {Catanzaro}, {Leone}, {Momany}, {Piotto} \&
  {D'Orazi}}{{Carretta} et~al.}{2010}]{Carretta2010}
{Carretta} E.,  {Bragaglia} A.,  {Gratton} R.~G.,  {Lucatello} S.,
  {Bellazzini} M.,  {Catanzaro} G.,  {Leone} F.,  {Momany} Y.,  {Piotto} G.,
  {D'Orazi} V.,  2010, \aap, 520, A95

\bibitem[\protect\citeauthoryear{{Carretta}, {Bragaglia}, {Gratton},
  {Lucatello}, {Catanzaro}, {Leone}, {Bellazzini}, {Claudi}, {D'Orazi},
  {Momany}, {Ortolani}, {Pancino}, {Piotto}, {Recio-Blanco} \&
  {Sabbi}}{{Carretta} et~al.}{2009}]{Carretta2009}
{Carretta} E.,  {Bragaglia} A.,  {Gratton} R.~G.,  {Lucatello} S.,  {Catanzaro}
  G.,  {Leone} F.,  {Bellazzini} M.,  {Claudi} R.,  {D'Orazi} V.,  {Momany} Y.,
   {Ortolani} S.,  {Pancino} E.,  {Piotto} G.,  {Recio-Blanco} A.,    {Sabbi}
  E.,  2009, \aap, 505, 117

\bibitem[\protect\citeauthoryear{{Castelli} \& {Kurucz}}{{Castelli} \&
  {Kurucz}}{2003}]{Castelli2003}
{Castelli} F.,  {Kurucz} R.~L.,  2003, in {Piskunov} N.,  {Weiss} W.~W.,
  {Gray} D.~F.,  eds, {Modelling of Stellar Atmospheres} Vol.~210 of {IAU
  Symposium}, {New Grids of ATLAS9 Model Atmospheres}.
p.~20P

\bibitem[\protect\citeauthoryear{{Cohen} \& {Kirby}}{{Cohen} \&
  {Kirby}}{2012}]{Cohen2012}
{Cohen} J.~G.,  {Kirby} E.~N.,  2012, \apj, 760, 86

\bibitem[\protect\citeauthoryear{{Cohen}, {Kirby}, {Simon} \& {Geha}}{{Cohen}
  et~al.}{2010}]{Cohen2010}
{Cohen} J.~G.,  {Kirby} E.~N.,  {Simon} J.~D.,    {Geha} M.,  2010, \apj, 725,
  288

\bibitem[\protect\citeauthoryear{{D'Antona}, {D'Ercole}, {Marino}, {Milone},
  {Ventura} \& {Vesperini}}{{D'Antona} et~al.}{2011}]{DAntona2011}
{D'Antona} F.,  {D'Ercole} A.,  {Marino} A.~F.,  {Milone} A.~P.,  {Ventura} P.,
     {Vesperini} E.,  2011, \apj, 736, 5

\bibitem[\protect\citeauthoryear{{Decin}, {Shkedy}, {Molenberghs}, {Aerts} \&
  {Aerts}}{{Decin} et~al.}{2004}]{Decin2004}
{Decin} L.,  {Shkedy} Z.,  {Molenberghs} G.,  {Aerts} M.,    {Aerts} C.,  2004,
  \aap, 421, 281

\bibitem[\protect\citeauthoryear{{Dupree}, {Strader} \& {Smith}}{{Dupree}
  et~al.}{2011}]{Dupree2011}
{Dupree} A.~K.,  {Strader} J.,    {Smith} G.~H.,  2011, \apj, 728, 155

\bibitem[\protect\citeauthoryear{{Federici}, {Bellazzini}, {Galleti}, {Fusi
  Pecci}, {Buzzoni} \& {Parmeggiani}}{{Federici} et~al.}{2007}]{Federici2007}
{Federici} L.,  {Bellazzini} M.,  {Galleti} S.,  {Fusi Pecci} F.,  {Buzzoni}
  A.,    {Parmeggiani} G.,  2007, \aap, 473, 429

\bibitem[\protect\citeauthoryear{{Freeman} \& {Rodgers}}{{Freeman} \&
  {Rodgers}}{1975}]{Freeman1975}
{Freeman} K.~C.,  {Rodgers} A.~W.,  1975, \apjl, 201, L71

\bibitem[\protect\citeauthoryear{{Harbeck}, {Smith} \& {Grebel}}{{Harbeck}
  et~al.}{2003}]{Harbeck2003}
{Harbeck} D.,  {Smith} G.~H.,    {Grebel} E.~K.,  2003, \aj, 125, 197

\bibitem[\protect\citeauthoryear{{Harris}}{{Harris}}{1996}]{Harris1996}
{Harris} W.~E.,  1996, \aj, 112, 1487

\bibitem[\protect\citeauthoryear{{Herwig}, {VandenBerg}, {Navarro}, {Ferguson}
  \& {Paxton}}{{Herwig} et~al.}{2012}]{Herwig2012}
{Herwig} F.,  {VandenBerg} D.~A.,  {Navarro} J.~F.,  {Ferguson} J.,    {Paxton}
  B.,  2012, \apj, 757, 132

\bibitem[\protect\citeauthoryear{{Hinkle}, {Wallace}, {Valenti} \&
  {Harmer}}{{Hinkle} et~al.}{2000}]{Hinkle2000}
{Hinkle} K.,  {Wallace} L.,  {Valenti} J.,    {Harmer} D.,  2000, {Visible and
  Near Infrared Atlas of the Arcturus Spectrum 3727-9300 A}.
Astronomical Society of the Pacific

\bibitem[\protect\citeauthoryear{{Icke} \& {Alca{\'{\i}}no}}{{Icke} \&
  {Alca{\'{\i}}no}}{1988}]{Icke1988}
{Icke} V.,  {Alca{\'{\i}}no} G.,  1988, \aap, 204, 115

\bibitem[\protect\citeauthoryear{{Johnson} \& {Pilachowski}}{{Johnson} \&
  {Pilachowski}}{2010}]{Johnson2010}
{Johnson} C.~I.,  {Pilachowski} C.~A.,  2010, \apj, 722, 1373

\bibitem[\protect\citeauthoryear{{Lee}, {Joo}, {Sohn}, {Rey}, {Lee} \&
  {Walker}}{{Lee} et~al.}{1999}]{Lee1999}
{Lee} Y.-W.,  {Joo} J.-M.,  {Sohn} Y.-J.,  {Rey} S.-C.,  {Lee} H.-C.,
  {Walker} A.~R.,  1999, \nat, 402, 55

\bibitem[\protect\citeauthoryear{{Mandushev}, {Staneva} \&
  {Spasova}}{{Mandushev} et~al.}{1991}]{Mandushev1991}
{Mandushev} G.,  {Staneva} A.,    {Spasova} N.,  1991, \aap, 252, 94

\bibitem[\protect\citeauthoryear{{Marino}, {Milone}, {Piotto}, {Cassisi},
  {D'Antona}, {Anderson}, {Aparicio}, {Bedin}, {Renzini} \&
  {Villanova}}{{Marino} et~al.}{2012}]{Marino2012}
{Marino} A.~F.,  {Milone} A.~P.,  {Piotto} G.,  {Cassisi} S.,  {D'Antona} F.,
  {Anderson} J.,  {Aparicio} A.,  {Bedin} L.~R.,  {Renzini} A.,    {Villanova}
  S.,  2012, \apj, 746, 14

\bibitem[\protect\citeauthoryear{{Marino}, {Milone}, {Piotto}, {Villanova},
  {Gratton}, {D'Antona}, {Anderson}, {Bedin}, {Bellini}, {Cassisi}, {Geisler},
  {Renzini} \& {Zoccali}}{{Marino} et~al.}{2011}]{Marino2011}
{Marino} A.~F.,  {Milone} A.~P.,  {Piotto} G.,  {Villanova} S.,  {Gratton}
  R.~G.,  {D'Antona} F.,  {Anderson} J.,  {Bedin} L.~R.,  {Bellini} A.,
  {Cassisi} S.,  {Geisler} D.,  {Renzini} A.,    {Zoccali} M.,  2011, \apj,
  731, 64

\bibitem[\protect\citeauthoryear{{Mateo}}{{Mateo}}{1998}]{Mateo1998}
{Mateo} M.~L.,  1998, \araa, 36, 435

\bibitem[\protect\citeauthoryear{{Mucciarelli}, {Bellazzini}, {Ibata}, {Merle},
  {Chapman}, {Dalessandro} \& {Sollima}}{{Mucciarelli}
  et~al.}{2012}]{Mucciarelli2012}
{Mucciarelli} A.,  {Bellazzini} M.,  {Ibata} R.~A.,  {Merle} T.,  {Chapman}
  S.~C.,  {Dalessandro} E.,    {Sollima} A.,  2012, \mnras, 426, 2889

\bibitem[\protect\citeauthoryear{{Norris}}{{Norris}}{2004}]{Norris2004}
{Norris} J.~E.,  2004, \apjl, 612, L25

\bibitem[\protect\citeauthoryear{{Norris}}{{Norris}}{2012}]{Norris2012}
{Norris} J.~E.,  2012, {CN, CH region line list}, private communication

\bibitem[\protect\citeauthoryear{{Pancino}, {Ferraro}, {Bellazzini}, {Piotto}
  \& {Zoccali}}{{Pancino} et~al.}{2000}]{Pancino2000}
{Pancino} E.,  {Ferraro} F.~R.,  {Bellazzini} M.,  {Piotto} G.,    {Zoccali}
  M.,  2000, \apjl, 534, L83

\bibitem[\protect\citeauthoryear{{Piotto}, {Villanova}, {Bedin}, {Gratton},
  {Cassisi}, {Momany}, {Recio-Blanco}, {Lucatello}, {Anderson}, {King},
  {Pietrinferni} \& {Carraro}}{{Piotto} et~al.}{2005}]{Piotto2005}
{Piotto} G.,  {Villanova} S.,  {Bedin} L.~R.,  {Gratton} R.~G.,  {Cassisi} S.,
  {Momany} Y.,  {Recio-Blanco} A.,  {Lucatello} S.,  {Anderson} J.,  {King}
  I.~R.,  {Pietrinferni} A.,    {Carraro} G.,  2005, \apj, 621, 777

\bibitem[\protect\citeauthoryear{{Simpson}, {Cottrell} \& {Worley}}{{Simpson}
  et~al.}{2012}]{Simpson2012}
{Simpson} J.~D.,  {Cottrell} P.~L.,    {Worley} C.~C.,  2012, \mnras, 427, 1153

\bibitem[\protect\citeauthoryear{{Skrutskie} et~al.,}{{Skrutskie}
  et~al.}{2006}]{Skrutskie2006}
{Skrutskie} M.~F.,  et~al., 2006, \aj, 131, 1163

\bibitem[\protect\citeauthoryear{{Smith}, {Suntzeff}, {Cunha}, {Gallino},
  {Busso}, {Lambert} \& {Straniero}}{{Smith} et~al.}{2000}]{Smith2000}
{Smith} V.~V.,  {Suntzeff} N.~B.,  {Cunha} K.,  {Gallino} R.,  {Busso} M.,
  {Lambert} D.~L.,    {Straniero} O.,  2000, \aj, 119, 1239

\bibitem[\protect\citeauthoryear{{Sneden}}{{Sneden}}{1973}]{Sneden1973}
{Sneden} C.~A.,  1973, PhD thesis, The University of Texas at Austin

\bibitem[\protect\citeauthoryear{{Stanford}, {Da Costa} \& {Norris}}{{Stanford}
  et~al.}{2010}]{Stanford2010}
{Stanford} L.~M.,  {Da Costa} G.~S.,    {Norris} J.~E.,  2010, \apj, 714, 1001

\bibitem[\protect\citeauthoryear{{Valcarce} \& {Catelan}}{{Valcarce} \&
  {Catelan}}{2011}]{Valcarce2011}
{Valcarce} A.~A.~R.,  {Catelan} M.,  2011, \aap, 533, A120

\bibitem[\protect\citeauthoryear{{van de Ven}, {van den Bosch}, {Verolme} \&
  {de Zeeuw}}{{van de Ven} et~al.}{2006}]{vandeVen2006}
{van de Ven} G.,  {van den Bosch} R.~C.~E.,  {Verolme} E.~K.,    {de Zeeuw}
  P.~T.,  2006, \aap, 445, 513

\bibitem[\protect\citeauthoryear{{van Leeuwen}, {Le Poole}, {Reijns}, {Freeman}
  \& {de Zeeuw}}{{van Leeuwen} et~al.}{2000}]{vanLeeuwen2000}
{van Leeuwen} F.,  {Le Poole} R.~S.,  {Reijns} R.~A.,  {Freeman} K.~C.,    {de
  Zeeuw} P.~T.,  2000, \aap, 360, 472

\bibitem[\protect\citeauthoryear{{van Loon}, {van Leeuwen}, {Smalley}, {Smith},
  {Lyons}, {McDonald} \& {Boyer}}{{van Loon} et~al.}{2007}]{vanLoon2007}
{van Loon} J.~T.,  {van Leeuwen} F.,  {Smalley} B.,  {Smith} A.~W.,  {Lyons}
  N.~A.,  {McDonald} I.,    {Boyer} M.~L.,  2007, \mnras, 382, 1353

\bibitem[\protect\citeauthoryear{{Villanova}, {Carraro}, {Scarpa} \&
  {Marconi}}{{Villanova} et~al.}{2010}]{Villanova2010}
{Villanova} S.,  {Carraro} G.,  {Scarpa} R.,    {Marconi} G.,  2010, \na, 15,
  520

\bibitem[\protect\citeauthoryear{{Villanova}, {Piotto}, {King}, {Anderson},
  {Bedin}, {Gratton}, {Cassisi}, {Momany}, {Bellini}, {Cool}, {Recio-Blanco} \&
  {Renzini}}{{Villanova} et~al.}{2007}]{Villanova2007}
{Villanova} S.,  {Piotto} G.,  {King} I.~R.,  {Anderson} J.,  {Bedin} L.~R.,
  {Gratton} R.~G.,  {Cassisi} S.,  {Momany} Y.,  {Bellini} A.,  {Cool} A.~M.,
  {Recio-Blanco} A.,    {Renzini} A.,  2007, \apj, 663, 296

\bibitem[\protect\citeauthoryear{{Worley} \& {Cottrell}}{{Worley} \&
  {Cottrell}}{2012}]{Worley2012}
{Worley} C.~C.,  {Cottrell} P.~L.,  2012, \pasa, 29, 29

\end{thebibliography}

\bsp

\label{lastpage}

\end{document}